\documentclass[10pt,conference]{IEEEtran}
\usepackage{fancyhdr, epsfig, epsf, amsthm, amsmath, amssymb, amsfonts, subfigure,color,dsfont}

\usepackage{cite}
\usepackage{graphicx}
\usepackage{dblfloatfix} 
\usepackage[most]{tcolorbox}
\tcbuselibrary{breakable} 
\usepackage{enumitem}
\usepackage[keeplastbox]{flushend}

\def\uav{u}
\def\nn{\nonumber}
\def\l{\left}
\def\r{\right}
\def\({\l(}
\def\){\r)}
\def\[{\l[}
\def\]{\r]}
\def\T{\intercal}

\IEEEoverridecommandlockouts

\begin{document}

\def\papertitle{Massive Autonomous UAV Path Planning: \\A Neural Network Based Mean-Field Game Theoretic Approach}

\title{ \fontsize{19}{26}\selectfont  \papertitle}

\author{\IEEEauthorblockN {Hamid Shiri, Jihong~Park, and Mehdi~Bennis} \vspace{-10pt}
\IEEEauthorblockA{\\\small Centre for Wireless Communications, University of Oulu, Finland, Email: \{hamid.shiri, jihong.park, mehdi.bennis\}@oulu.fi} 
\vspace{-.7cm}
}

\maketitle \thispagestyle{empty}

\begin{abstract}
This paper investigates the autonomous control of massive unmanned aerial vehicles (UAVs) for mission-critical applications (e.g., dispatching many UAVs from a source to a destination for firefighting). Achieving their fast travel and low motion energy without inter-UAV collision under wind perturbation is a daunting control task, which incurs huge communication energy for exchanging UAV states in real time. We tackle this problem by exploiting a mean-field game (MFG) theoretic control method that requires the UAV state exchanges only once at the initial source. Afterwards, each UAV can control its acceleration by locally solving two partial differential equations (PDEs), known as the Hamilton-Jacobi-Bellman (HJB) and Fokker-Planck-Kolmogorov (FPK) equations. This approach, however, brings about huge computation energy for solving the PDEs, particularly under multi-dimensional UAV states. We address this issue by utilizing a machine learning (ML) method where two separate ML models approximate the solutions of the HJB and FPK equations. These ML models are trained and exploited using an online gradient descent method with low computational complexity. Numerical evaluations validate that the proposed ML aided MFG theoretic algorithm, referred to as \emph{MFG learning control}, is effective in collision avoidance with low communication energy and acceptable computation energy.
\end{abstract}


\section{Introduction}

Many unmanned aerial vehicles (UAVs) are essential in mission-critical applications, for covering wide disaster sites in emergency cellular networks~\cite{b3} and for delivering heavy payload in rescue mission and firefighting scenarios~\cite{Ackerman:18, b45}. These applications are real-time, and do not tolerate remote control delays from a central controller. Besides, they necessitate reliable control under uncertainty such as wind perturbations, making pre-programed offline control algorithms ill-suited. In view of this, in this paper we focus on the problem of controlling a large number of UAVs in a distributed and online way, so as to achieve 1) the \emph{fastest travel} from a source to a destination, while jointly minimizing 2)~\emph{motion energy}, and 3)~\emph{inter-UAV collision}, under wind dynamics.

\begin{figure}[t]
\centering
\includegraphics[width = \columnwidth]{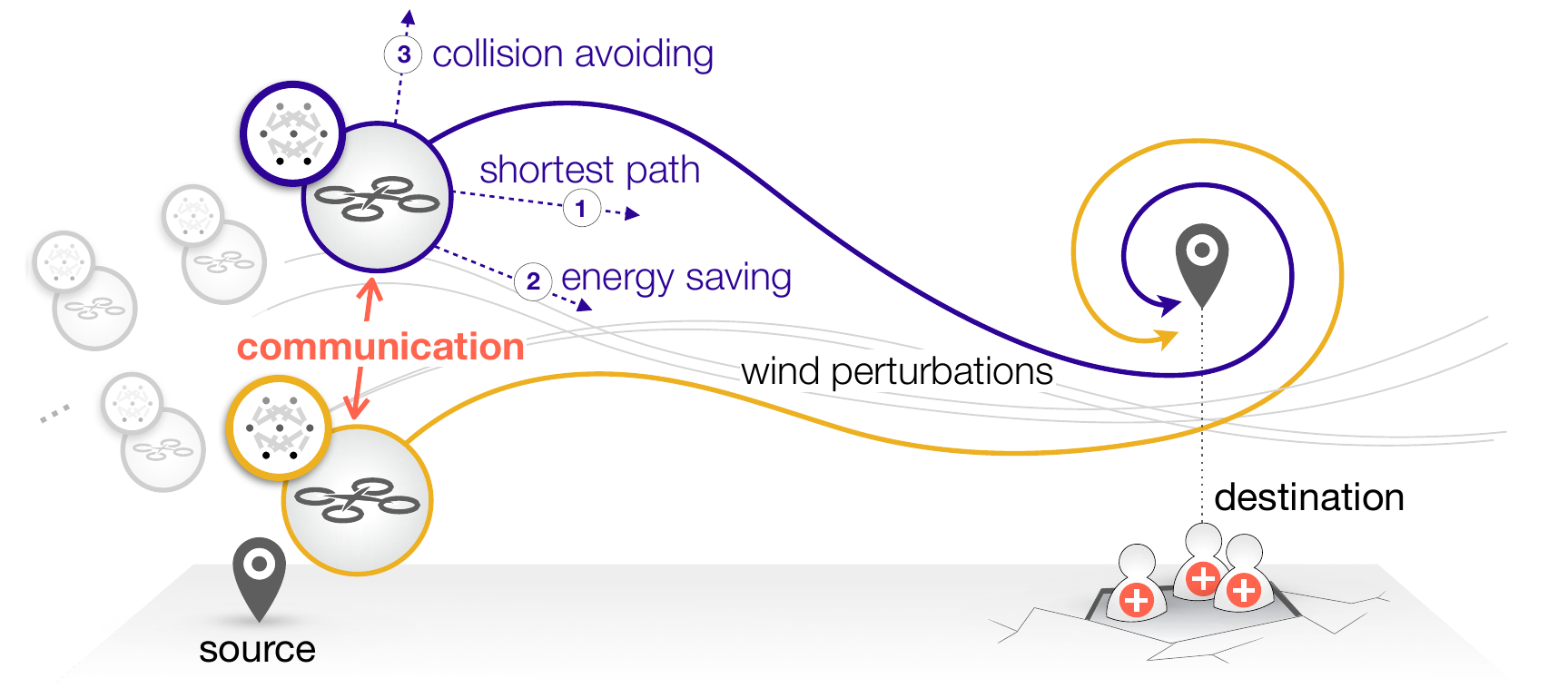}\vskip -5pt
\caption{An illustration of dispatching massive UAVs from a source point to a destination site. Each UAV communicates with neighboring UAVs for achieving: 1) the fastest travel, while jointly minimizing 2) motion energy and 3) inter-UAV collision, under wind perturbations.} \vskip -5pt
\label{Fig:Overview}
\end{figure}

This problem is challenging as illustrated in Fig.~\ref{Fig:Overview}, wherein each UAV is faced with making control decisions with many degrees of freedom, while taking into account energy-saving and collision-avoidance. For collision avoidance, multiple UAVs need to interact with each other, which require inter-UAV communications whose delay and/or energy cost increases exponentially with the number of UAVs. Such communication and control overhead is persistent as the control must be continual under wind perturbations.

To address the aforementioned issues, we leverage mean-field game (MFG) theory~\cite{b11, b12}, a mathematical framework that is effective in reducing the communication and control overhead of distributed control under agent interactions (e.g., collisions) through their states (e.g., locations)~\cite{b3}. At its core, MFG considers a large number of agents, each of which approximately views the other agents' states as the \emph{global state} averaged across all agents. The global state is identically given for all agents at any given time, and one can thus focus only on controlling a single agent while incorporating its interactions via the global state distribution, referred to as the \emph{mean-field (MF) distribution}. 

The said MFG theoretic control is operated by locally solving two partial differential equations (PDEs) at each agent. Namely, a single agent computes the MF distribution by solving the \emph{Fokker-Plank-Kolmogorov (FPK)} equation, so long as the initial global state is known by exchanging agents' states only once. For the given MF distribution, the optimal control of the agent is determined by solving the other PDE induced by a continuous-time Markov decision problem (MDP), known as the \emph{Hamilton-Jacobi-Bellman (HJB)} equation \cite{b12}.

While effective, MFG theoretic approaches are computationally expensive due to solving both HJB and FPK equations, particularly with multi-dimensional states~\cite{b50}, limiting their adoption for real-time multi-dimensional control applications. To circumvent this problem, we propose an \emph{MFG learning control} algorithm in which the HJB and FPK solutions are approximated using two separate machine learning (ML) models (e.g., neural networks), denoted as \emph{HJB model} and \emph{FPK model}, respectively. The HJB and FPK models stored at each UAV are simultaneously trained and exploited for control in an online manner. Numerical evaluations validate that the proposed MFG learning control more reliably guarantees collision avoidance with significant communication energy reduction, at the cost of a slight increase in computation and motion energy consumption.

\vspace{5pt}\noindent \textbf{Related works}.\quad
The problem of UAV placement for supporting communication systems has been studied in \cite{b40}. Under wind perturbations, the real-time placement of massive UAVs without collision has been investigated in \cite{b3}. Path planning is a more challenging problem wherein UAVs are controlled to reach a destination. In offline control, a multiple-UAV scenario have been addressed in \cite{b40}. In online control, an evolutionary algorithm \cite{b46} and a partially observable Markov decision process based method \cite{b47} have been proposed. For other communication and control related issues for UAV systems, readers are encourage to check \cite{b48} and \cite{b45}, respectively.

\section{System model} \label{SE:02}
We assume a set $ \mathcal{N} $ of $ N $ UAVs traveling from a common source to a destination in a two-dimensional plane, where the origin is set as the destination. At time $t\geq 0$, the $i$-th UAV $\uav_i\in\mathcal{N}$ controls its acceleration~$a_i(t)\in\mathbb{R}^2$, so as to minimize its: \textbf{1) travel time}, \textbf{2) motion energy}, and \textbf{3) inter-UAV collision}, during the remaining travel to the destination.

The control $a_i(t)$ of $\uav_i$ is based not only on its local state~$s_i(t)$, but also on the states $s_{-i}(t)$ of a set $\mathcal{N}_i(t)\!\subset\!\mathcal{N}$ of the other $(N_i-1)$ UAVs within $\uav_i$'s communication range $d>0$ with $N_i(t) = |\mathcal{N}_i(t)|\leq N$ for collision avoidance, and $s_{-i}(t) \!=\! \{s_j(t)\! \mid \!\|r_j(t)\!-\!r_i(t)\|\!\leq\! d, \forall j \neq i \}$. The communication range $d$ is determined by the minimum received signal-to-noise ratio ($\textsf{SNR}$) $\theta>0$ required for successful decoding under the standard path loss model, which is given as $d = [P/(\theta\sigma^2)]^{1/\alpha}$ with an identical transmission power $P$, noise power $\sigma^2$, and path loss exponent $\alpha\geq 2$. 

The state~$s_i(t)= [r_i(t)^{\T},v_i(t)^{\T}]^{\T}\in\mathbb{R}^4$ of UAV $\uav_i$ is comprised of its location $r_i(t)\in\mathbb{R}^2$ and velocity $v_i(t)\in\mathbb{R}^2$ that are dynamically updated by the control $a_i(t)$ under random wind dynamics. Following \cite{b49}, the wind dynamics are assumed to follow an Ornstein-Uhlenbeck process with an average wind velocity~$v_o$. The temporal state dynamics are thereby given~as:

\vspace{-10pt}\small\begin{align} 
\text{d}v_i(t) &= a_i(t) \text{d}t - c_0 \left ( v_i(t) - v_o \right ) \text{d}t + V_o \text{d}W_i(t) \label{Eq:dv}\\
\text{d}r_i(t) &= v_i(t)\text{d}t, \label{Eq:dr}
\end{align}\normalsize
where $ c_0 $ is a positive constant, $ V_o\in\mathbb{R}^{2\times 2}$ is the covariance matrix of the wind velocity, and $W_i(t)\in\mathbb{R}^2 $ is the standard Wiener process independently and identically distributed (i.i.d.) across UAVs. 

To achieve the aforementioned goals \textbf{1)}, \textbf{2)}, and \textbf{3)}, UAV~$\uav_i$ at time $t\!<\!T$ aims to minimize its average cost~$\psi_i(t)$, where the average is taken with respect to the measure induced by all possible controls for $\tau\!\in\![t,T]$. The cost $\psi_i(t)$ consists of the term $\phi_L\!\(s_i(t)\)$ depending only on the local state $s_i(t)$ and the term $\phi_G\!\(s_{N_i}\!(t)\)$ relying on the global state $s_{N_i}\!(t) = [s_i(t)^\T, s_{-i}(t)^\T]^\T \in \mathbb{R}^{4 N_{i}}$ observed by $\uav_i$, given~as:

\vspace{-10pt}\small\begin{align} \label{Eq:LRA}
\hspace{-5pt} \psi_i(t) \!=\!  \mathsf{E} \[\; \int_{t}^{\T}  \bigg( 
\phi_L\!\(s_i(\tau)\) + c_4 \phi_G\!\(s(\tau)\)
\bigg) \textup{d}\tau \]
\end{align}\normalsize
where
\vspace{-10pt}\small\begin{align}
&\;\;\phi_L\!\big(s_i(t)\big) = \overbrace{\frac{v_i(t) \cdot r_i(t)}{\left \| r_i(t) \right \|} +  c_1 {\left\| r_i(t) \right\|^{2}}}^{\text{\textbf{1) travel time} minimization}}  + \overbrace{c_2 {\left\| v_i(t) \right\|^{2}} + c_3 {\left\| a_i(t) \right\|^{2}  } }^{\text{\textbf{2) motion energy} minimization}}, \nn\\
&\;\;\phi_G\!\big(s_{N_i}\!(t)\big)  =  \underbrace{\frac{1}{N_i(t)-1} \sum_{\uav_j\in\mathcal{N}_i(t)\backslash \{\uav_i\}}   \frac{ \left\| v_{j}(t) - v_i(t) \right\|^{2} }{\left( \varepsilon + \left \| r_j(t) - r_i(t) \right \|^2  \right)^{\beta}}}_{\text{\textbf{3) collision} avoidance \& connectivity guarantee}}\nn,
\end{align}\normalsize
and the terms $ c_1 $, $ c_2 $, $ c_3 $, $\beta$, and $\varepsilon$ are positive constants. 

\begin{figure*}[b]
\hrulefill
\setcounter{equation}{5}
\footnotesize\begin{align} 
\label{Eq:HJB}
\hspace{-43pt}\text{(For \textbf{HJB})}\hspace{5pt}\mathsf{H}\big(\psi_i\!(t);s_{N_i}\!(t) \big)& = \partial_{t} \psi_i\!(t) + \inf_{a_i(t)} \bigg\{ \big[ A s_i(t) + B (a_i(t) + c_0 v_o ) \big]^{\T}  \nabla \psi_i\!(t)
+  \frac{1}{2}\textup{tr}\!\(G G^{\T} \nabla^2 \psi_i\!(t)\) + \phi_L\!\(s_i(t)\) + \phi_G\!\(s_{N_i}\!(t)\)   \bigg\}\\
\label{Eq:HJBmodified}
&= \partial_{t} \psi_i\!(t)  +   \[ A s_i(t)   -  \frac{1}{4 c_3} B B^{\T} \nabla \psi_i\!(t)  +  c_0 v_o B   \]^{\T}  \nabla \psi_i\!(t)  
+ \frac{1}{2}\textup{tr}\!\(G G^{\T} \nabla^2 \psi_i\!(t)\) +   \phi_L\!\(s_i(t)\) + \phi_G\!\(s_{N_i}\!(t)\) \end{align}
\hrulefill\setcounter{equation}{7}
\begin{align}
\label{Eq:HJBmf}
\hspace{-8pt}\text{(For \textbf{MFG})}\hspace{4pt}\mathsf{H}\big(\psi_i\!(t);s_i(t),m(t) \big) &\!=\! \partial_{t} \psi_i\!(t)  +   \[ A s_i(t)   -  \frac{1}{4 c_3} B B^{\T} \nabla \psi_i\!(t)  +  c_0 v_o B  \]^{\T}  \!\nabla \psi_i\!(t)  
+ \frac{1}{2}\textup{tr}\!\(G G^{\T} \nabla^2 \psi_i\!(t)\) +   \phi_L\!\(s_i(t)\) + \phi_G\!\(s_i(t),m(t)\) \\
\label{Eq:FPK}
\mathsf{F}\!\big(m(t); s_i(t), \psi_i\!(t) \big) & \!=\! \partial_{t} m(t) + \nabla\!\bigg(\bigg[ A s_i(t) + B \(a_i^*(t) + c_0 v_o \) \bigg] m(t)\bigg) 
- \frac{1}{2}\textup{tr}\!\(G G^{\T} \nabla^2  m(t)\)\\
\label{Eq:FPKmodified}
&\!=\! \partial_{t} m(t) + \nabla\!\(\[ A s -  \frac{1}{2 c_3} B B^{\T} \nabla \psi_i\!(t)  +  c_0 v_o B  \] m(t)\)
-\frac{1}{2}\textup{tr}\!\(G G^{\T} \nabla^2  m(t)\)
\end{align}\normalsize
\end{figure*}
\setcounter{equation}{3}

The local term $\phi_L\!\(s_i(t)\)$ in \eqref{Eq:LRA} focuses on the following two objectives. For \textbf{1) travel time} minimization, it is intended to minimize the remaining travel distance $\|r_i(t)\|^2$, while maximizing the velocity towards the destination, i.e., minimizing the projected velocity $v_i(t)\cdot r_i(t)/\|r_i(t)\|$ towards the opposite direction to the destination. For \textbf{2) motion energy} minimization, it is planned to minimize the kinetic energy and the acceleration control energy that are proportional to $\|v_i(t)\|^2$ and $\| a_i(t) \|^2$, respectively~\cite{b35, b6}.

The global term $\phi_G\!\(s_{N_i}\!(t)\)$ in \eqref{Eq:LRA} refers to \textbf{3)~collision} avoidance, and is intended to form a flock of UAVs moving together~\cite{b1}. The flocking leads to small relative inter-UAV velocities for avoiding collision even when their controlled velocities are slightly perturbed by wind dynamics. Furthermore, the flocking yields closer inter-UAV distances without collision. This is beneficial for allowing more UAVs to exchange their states, i.e., larger $N_i(t)$, thereby contributing also to collision avoidance. In view of this, we adopt the Cucker-Smale flocking~\cite{b3,b1} that reduces the relative velocities for the UAVs. The relative velocity $\|v_j(t)-v_i(t)\|$ and the inter-UAV distance $\|r_j(t)-r_i(t)\|$ are thus incorporated in the numerator and denominator of $\phi_G\!\(s_{N_i}\!(t)\)$, respectively.
 
Incorporating the cost function \eqref{Eq:LRA} under its temporal dynamics \eqref{Eq:dv} and \eqref{Eq:dr}, the control problem of UAV $\uav_i$ at time~$t$ is formulated as:
 
\vspace{-10pt}\small\begin{align} \label{Eq:psi}
&\psi_i^*\!(t) = \min_{a_i(t)}\; \psi_i(t)  \\
\text{s.t.}\quad &\text{d} s_i(t)= \left( A s_i(t) + B (a_i(t) + c_0 v_o ) \right) \textup{d}t + G \text{d} W_i(t), \label{Eq:StateDyn}
\end{align}\normalsize
where $A\!=\!\(\begin{smallmatrix}0 & I\\0 & -c_0 I\end{smallmatrix}\)$, $B\!=\!\(\begin{smallmatrix}0 \\ 
I\end{smallmatrix}\)$, $G\!=\!\(\begin{smallmatrix}0 \\ V_o\end{smallmatrix}\)$, and $I$ denotes the two-dimensional identity matrix. The minimum cost $\psi_i^*\!(t)$ is referred to as the \emph{value function} of the optimal control, and is derived using two different control methods in the next section.

\begin{figure}
	\centering
	\subfigure[HJB (learning) control.]{\includegraphics[width= .385\columnwidth]{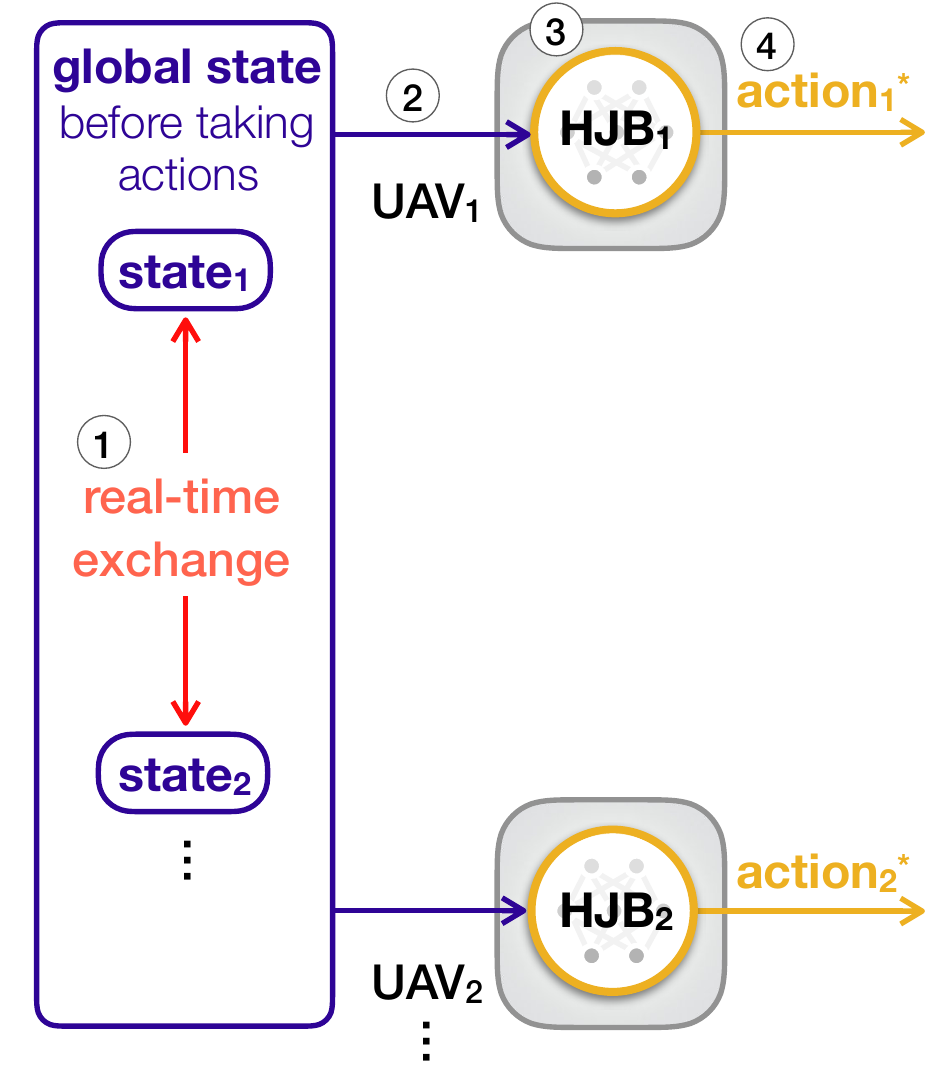}}\hspace{10pt} 
	\subfigure[MFG (learning) control.]{\includegraphics[width=  .51\columnwidth]{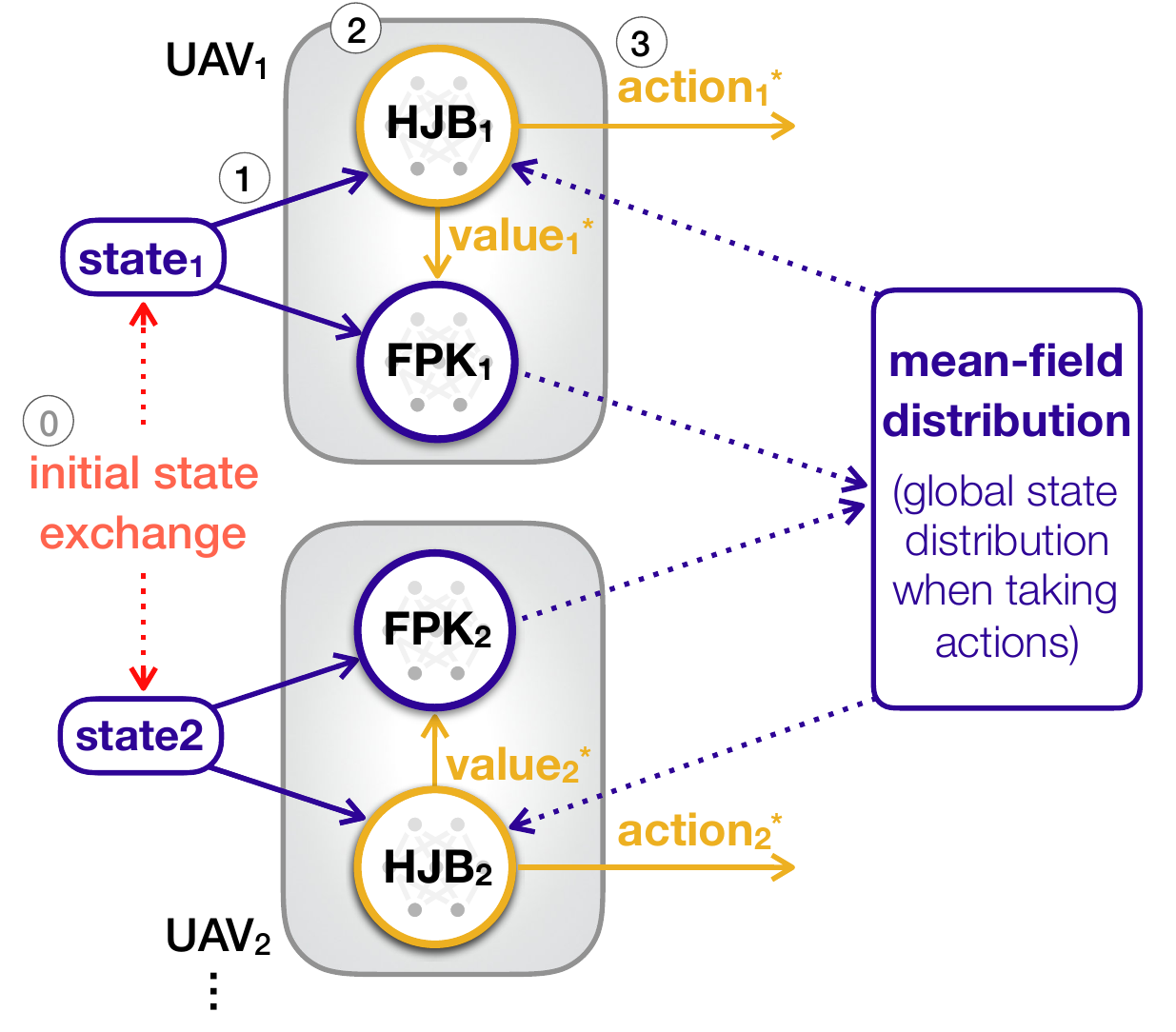}} \vskip -7pt
	\caption{\small Operational structures of HJB (learning) control and MFG (learning) control.}\label{Fig:HJBMFG} \vskip -13pt
\end{figure}

\section{HJB Control and MFG Control} \label{SE:03}
Deriving the UAV $\uav_i$'s value function $\psi_i^*(t)$ in \eqref{Eq:psi} is intertwined with other UAVs, through the collision avoidance term $\phi_G\!\(s_{N_i}\!(t)\)$ in \eqref{Eq:LRA}. Therefore, this is an $N_i$-player non-cooperative game whose well-known solution is the Nash equilibrium (NE), i.e., the control decisions under which no UAV can unilaterally decrease its cost~\cite{b12}. Its solution complexity exponentially increases with $N_i$, which is a poor fit for real-time applications. 
To address this pressing concern, in this section we consider two different control methods: \emph{1) HJB control}, our baseline method in which each UAV's control only takes into account the other UAVs' \emph{states before taking their actions}; and \emph{2) MFG control}, our proposed method that incorporates the intertwined controls via an approximated global state distribution, i.e., the \emph{MF distribution}. 

It is noted that HJB control does not always achieve the NE as it intentionally neglects the actual control interactions, i.e., the \emph{states when taking actions}. On the other hand, MFG control relies on the MF approximation, and only achieves the NE asymptotically when $N\rightarrow \infty$~\cite{b12}. The operational details of both control schemes are elaborated in the following subsections, and their effectiveness under a large finite number of UAVs will be numerically examined in Sec.~\ref{SE:05}.

\subsection{HJB Control} \label{Sec:HJBcontrol}
The UAV $\uav_i$'s value function $\psi_{i}^*(t)$ in \eqref{Eq:psi} is equivalent to the solution of its corresponding HJB $\mathsf{H}\big(\psi_i^*\!(t);s_{N_i}\!(t) \big)\!=\!0$ formulated according~to the Markov decision principle. The left-hand side $\mathsf{H}\big(\psi_i^*\!(t);s_{N_i}\!(t) \big)$ is given by putting $\psi_{i}^*(t)$ into~$\mathsf{H}\big(\psi_i\!(t);s_{N_i}\!(t) \big)$ in~\eqref{Eq:HJBmodified} at the bottom of the next page (see the derivation details in~\cite{b12}). Due to the global term $\phi_G(s_{N_i}\!(t))$ therein for collision avoidance, the HJB solution requires collecting the other UAVs' states. Furthermore, achieving the NE of $N_i$-UAV controls, necessitates solving $N_i$-coupled HJBs whose required number of state exchanges exponentially increases with $N_i$. For example, each HJB is first solved while the other $(N_i-1)$ UAVs' states are fixed, and this should be iterated for $N_i$ UAVs in a recursive manner until all action changes stop, i.e., convergence to the NE~\cite{b12}. The said $N_i$-coupled HJB solutions require $N_i\times(N_i-1)\times K$ state exchanges per time instant $t$, where $K$ denotes the number of iterations until convergence to the NE.

Such excessive communication overhead is not bearable for real-time UAV controls. Therefore, while compromising convergence to the NE, as a baseline control scheme we instead consider HJB control of UAVs that exchange $\sum_{i=1}^{N_i} N_i(t)$ number of states before solving the HJBs, i.e., before taking actions, at each time instant~$t$. Afterwards, each HJB is solved independently without recursion, as visualized in Fig.~\ref{Fig:HJBMFG}-a. At time $t$, $\uav_i$'s HJB control is summarized as below.

\vspace{3pt}\begin{tcolorbox}[colback=white, colframe=blue!30!black, boxrule=.5pt, top=3pt, bottom=0pt,left=10pt, right=5pt, arc=0mm]
\hspace{-8pt}\noindent\textbf{Algorithm 1. HJB Control}
\begin{enumerate}[leftmargin=8pt,label={\arabic*)}]
\item Collect the states $s_{-i}(t)$ from $(N_i(t)-1)$ UAVs.

\item Calculate the value $\psi_i^*\!(t)$ by solving the HJB $\mathsf{H}\big(\psi_i^*\!(t);s_{N_i}\!(t) \big)=0$ (see \eqref{Eq:HJBmodified}).

\item Take the optimal action $a_i^*(t) \!=\! \frac{1}{2c_3}B^\T\nabla\psi_i^*\!(t)$.
\end{enumerate}
\end{tcolorbox}

Here, \eqref{Eq:HJBmodified} is derived by applying the optimal control $a_i^*(t)$ to \eqref{Eq:HJB}, where $\nabla$ denotes the differential operator taken with respect to $s_i(t)$. The optimal control $a_i^*(t)\!=\! \frac{1}{2c_3}B^\T\nabla\psi_i^*\!(t)$ is obtained according to the Karush-Kuhn-Tucker (KKT) conditions, since the HJB's Hamiltonian, i.e., the terms inside the infimum in \eqref{Eq:HJB}, is convex with respect to~$ a_i(t) $. The existence of $a_i^*(t)$ is ensured by the fact that the HJB with \eqref{Eq:HJB} has a unique solution $\psi_i^*\!(t)$ according to \cite{b12}, as long as the drift term $A s_i(t) + B (a_i(t) + c_0 v_o)$ in \eqref{Eq:StateDyn} and the instantaneous cost $\phi_L\!\(s_i(t)\) + \phi_G\!\(s_{N_i}\!(t)\)$ are smooth, i.e., continuous first derivatives.

\subsection{MFG Control} \label{Sec:MFGcontrol}
Compared to HJB control with $\sum_{i=1}^{N_i} N_i(t)$ state exchanges \emph{per time instant $t$}, MFG control requires $N\times (N-1)$  state exchanges \emph{only at the initial time $t=0$}, while asymptotically guaranteeing the NE anytime as $N$ goes to infinity. This is viable by locally calculating the MF distribution $m(t)$ that asymptotically converges to the (empirical) global state distribution when all actions are taken under the NE, i.e., $\lim_{N\to \infty}\!\frac{1}{N} \sum_{i=1}^N \mathds{1}_{s_i(t)}\!=\! m(t) $. With finite UAVs, it yields an MF approximation that achieves the $\epsilon$-NE~\cite{b12}.

To this end, each UAV under MFG control locally solves a pair of the HJB $\mathsf{H}\big(\psi_i^*\!(t);s_i(t),m(t)\big)=0$ (see \eqref{Eq:HJBmf} with $\psi_i^*\!(t)$) and its coupled FPK $\mathsf{F}\!\big(m(t); s_i(t), \psi_i^*\!(t) \big)=0$ (see \eqref{Eq:FPKmodified} with $\psi_i^*\!(t)$) that is derived from the state dynamics \eqref{Eq:StateDyn} with the It\^{o}'s lemma \cite{b12}. As illustrated in Fig.~\ref{Fig:HJBMFG}-b, solving the HJB produces the value $\psi_i^*\!(t)$ (or its corresponding optimal action $a_i^*(t)$), which is fed to the FPK whose solution is the MF distribution $m(t)$. This operation is locally iterated $K$ times until it converges to the NE. At time~$t$, $\uav_i$'s MFG control is described as follows.

\vspace{3pt}\begin{tcolorbox}[colback=white, colframe=blue!30!black, boxrule=.5pt, top=3pt, bottom=3pt,left=10pt, right=5pt, arc=0mm]
\hspace{-8pt}\noindent\textbf{Algorithm 2. MFG Control}

\hspace{-8pt}\noindent For $k\in[1,K]$:
\begin{enumerate}[leftmargin=8pt,label={\arabic*)}]
\item Calculate the value $\psi_i^{[k]}\!(t)$ by solving the HJB $\mathsf{H}\big(\psi_i^{[k]}\!(t);s_i(t),m^{[k-1]}\!(t)\big)=0$ (see \eqref{Eq:HJBmf}).

\item Calculate the MF distribution $m^{[k]}\!(t)$ by solving the FPK $\mathsf{F}\big(m^{[k]}\!(t); s_i(t), \psi_i^{[k]}\!(t) \big)$.

\item Iterate 1) and 2) until $k=K$.

\item Take the optimal action $a_i^*(t) \!=\! \frac{1}{2c_3}B^\T\nabla\psi_i^{[K]}\!(t)$.
\end{enumerate}
\vspace{-5pt}\hrulefill

\hspace{-8pt}\noindent Initial MF distribution $m^{[0]}\!(t)$ at $k=0$:
\begin{itemize}[leftmargin=8pt]
\item If $t=0$, $m^{[0]}\!(0)= 1/N\sum_{i=1}^N \mathds{1}_{s_i(t)}$, computed by collecting the states $s_{-i}(0)$ from N UAVs.
\item Otherwise, $m^{[0]}\!(t)=m^{[K-1]}\!(t-\Delta t)$, i.e., the converged MF distribution in the previous control where $\Delta t$ denotes the control interval.
\end{itemize}
\end{tcolorbox}

It is noted that the HJB's global term $\phi_G(s_i(t),m(t))$ in \eqref{Eq:HJBmf} approximates $\phi_G\!\(s_{N_i}\!(t)\)$ in \eqref{Eq:HJBmodified}, where

\setcounter{equation}{10}
\vspace{-10pt}\small\begin{align} \label{Eq:GlobalMFG}
\phi_G(s_i(t), m(t)) \!=\!  \int_s  m(t) \frac{  \left\|v(t) - v_i(t) \right\|^{2} }{\left( \varepsilon^2 + \left \| r(t) - r_i(t) \right \|)^2  \right)^\beta}  \text{d}s.
\end{align}\normalsize
This MF approximation is based on treating each of the UAVs' states as $s=[r(t)^{\T},v(t)^{\T}]^{\T}$ induced by the MF distribution $m(t)$. The approximation converges to the exact value as $N\rightarrow\infty$, so long as $\phi_G\!\(s_{N_i}\!(t)\)$ is bounded and UAV indices are permutable, i.e., the exchangeability of actions for the same states (see the condition details in~\cite{b12}).


\section{ML Aided HJB and MFG Controls} \label{SE:04}
Both HJB and MFG controls are facilitated by the HJB and FPK equations. These PDEs are solved by discretizing the domain in a way that the derivatives therein can be approximated using finite differences. Unfortunately, such a finite difference method requires finer discretization as the domain dimension increases, incurring higher computational complexity. For instance, in a two-dimensional $x$-$y$ domain, the convergence of a numerical PDE solution with the temporal discretization step size $\Delta t$ is guaranteed by the Courant-Friedrichs-Lewy (CFL) condition $\Delta t \leq (\Delta x^{-1} + \Delta y^{-1})^{-1}$ whose feasible step size is smaller than the required step size in a one-dimensional domain, i.e., $\Delta t \leq \Delta x$~\cite{b50}.

To enable multi-dimensional control in real time with low computational complexity, we propose \emph{HJB learning control} and \emph{MFG learning control} that approximate both HJB control and FPK control in Sec.~\ref{SE:03}, respectively. Via these methods, ML models learn to solve the HJB and FPK in an online way, as elaborated in the following subsections.

\subsection{HJB Learning Control} \label{SUBSE:04.01}
HJB learning control exploits ML to enable and represent the baseline method, HJB control in Sec.~\ref{Sec:HJBcontrol}. The key idea is to approximate the problem of solving the HJB equation $\mathsf{H}\big(\psi_i^*\!(t);s_{N_i}\!(t) \big)=0$ by minimizing $\mathsf{H}\big(\hat{\psi}_i\!(t);s_{N_i}\!(t) \big)$ via a data-driven regression method as proposed in \cite{b2}. To this end, a single hidden layer ML model, hereafter referred to as an \emph{HJB model}, is constructed at the UAV $\uav_i$. Its input $s_{N_i}\!(t)$ is fed to $M_\textsf{H}$ hidden nodes with a given activation function $\sigma_\textsf{H}(\cdot)$, which are fully connected to the model output $\hat{\psi}_i(t)$ through a weight vector $w_{i,\textsf{H}}(t)$, i.e., 

\vspace{-10pt}\small\begin{align}
\hat{\psi}_i(t) = w_{i,\textsf{H}}(t)^\T\sigma_\textsf{H}\!\(s_{N_i}\!(t)\).
\end{align} \normalsize

The model is trained by adjusting $w_{i,\textsf{H}}(t)$ per each observation $s_{N_i}\!(t)$, so as to minimize its cost function $L_{i,\textsf{H}}(t)$ comprising a loss function $\ell_{i,\textsf{H}}(t)$ and a regularizer $R_i(t)$: 

\vspace{-10pt}\small\begin{align} \label{Eq:HJBlearncost}
\hspace{-6pt}L_{i,\textsf{H}}(t) = \underbrace{\frac{1}{2} \l| \hat{\mathsf{H}}\big(\hat{\psi}_i(t);s_{N_i}\!(t)\big) \r|^{2}}_{\ell_{i,\textsf{H}}(t)} +\; c_{\hspace{.5pt}\textsf{H}}  \underbrace{\max\l\{0, s_i(t)^\T\frac{\text{d}s_i(t)}{\text{d}t}\r\} }_{R_i(t)},
\end{align}\normalsize \vskip -5pt
\noindent where $c_{\hspace{.5pt}\textsf{H}}$ is a positive constant. The loss function is intended to minimize $\hat{\mathsf{H}}\big(\hat{\psi}_i\!(t);s_{N_i}\!(t) \big)$ in \eqref{Eq:HJBmodified}. The regularizer is meant to stop the movement when reaching the destination, i.e., $s_i(T) = [r_i(T)^\T, v_i(T)^\T]^\T \!=\! 0$. At time $t$, $\uav_i$'s HJB learning control is given as below.

\vspace{3pt}\begin{tcolorbox}[colback=white, colframe=blue!30!black, boxrule=.5pt, top=3pt, bottom=0pt,left=10pt, right=0pt, arc=0mm] 
\hspace{-8pt}\noindent\textbf{Algorithm 3. HJB Learning Control}
\begin{enumerate}[leftmargin=8pt,label={\arabic*)}] 
\item Collect the states $s_{-i}(t)$ from $(N_i(t)-1)$ UAVs.

\item Update the weight $w_{i,\textsf{H}}(t)$ as:

\vspace{-10pt}\small\begin{align}
\hspace{-5pt} w_{i,\textsf{H}}(t) \!=\! w_{i,\psi}(t\!-\!\Delta t) \!-\! \mu \text{sign}\(\nabla_{\!w} \ell_{i,\textsf{H}}(t)\) \!-\! c_{\hspace{.5pt}\textsf{H}}\nabla_{\!w} R_i(t). \nn
\end{align}\normalsize  

\item Calculate the value $\hat{\psi}_i(t) = w_{i,\psi}(t)^\T \sigma_\psi(s_{N_i}\!(t))$.

\item Take the optimal action $a_i^*(t) \!=\! \frac{1}{2c_3}B^\T\nabla\hat{\psi}_i(t)$.
\end{enumerate}
\end{tcolorbox}
The weight update rule in 2) of Algorithm 3 is derived by applying a normalized gradient descent algorithm (NGD), modified from the gradient descent algorithm (GD) in order to avoid saddle points under non-convex loss functions \cite{b51}. To be specific, the weight update rule under GD with the step size $\mu>0$ is $w_{i,\textsf{H}}(t)=w_{i,\textsf{H}}(t-\Delta t) - \mu \nabla_{\!w}L_{i,\textsf{H}}(t)$, which is modified as $w_{i,\textsf{H}}(t)=w_{i,\textsf{H}}(t-\Delta t) - \mu \text{sign}\(\nabla_{\!w}L_{i,\textsf{H}}(t)\)$ under the original NGD~\cite{b51} with $\text{sign}(x) = x/\| x\|$. As opposed to this, the weight update rule in Algorithm 3 applies the sign operation only to the loss function $\ell_{i,\textsf{H}}(t)$ in $L_{i,\textsf{H}}(t)$, in order not to disturb $R_i(t)$ activations as detailed next.

The regularizer $R_i(t)$ aims to ensure stably reaching the destination without further movement, i.e., the terminal zero-state convergence $s_i(T)\!=\!0$. With this, $R_i(t)$ becomes activated, i.e., $R_i(t)\!>\!0$, for penalizing the loss function $\ell_{i,\textsf{H}}(t)$, when the state change direction (the sign of $\text{d}s_i(t)/\text{d}t$) under the current control is the same as the current state direction (the sign of $s_i(t)$), i.e., $s_i(t)^\T \text{d}s_i(t)/\text{d}t>0$. Otherwise, the current control is capable of stabilizing the state, and the regularizer is thus inactivated, i.e., $R_i(t)\!=\!0$. The regularizer activations during UAV travels will be discussed in Sec. V.

In the loss function $\ell_{i,\mathsf{H}}(t)$, the expression $\hat{\mathsf{H}}\big(\hat{\psi}_i(t);s_{N_i}\!(t)\big)$ is derived by applying $\hat{\psi}_i(t)$ to \eqref{Eq:HJBmodified} with the same procedure as described in Algorithm~1, except for the following detail. The cost function $L_{i,\textsf{H}}(t)$ includes ${\text{d}s_i(t)}/{\text{d}t}$; namely, within $R_i(t)$ in \eqref{Eq:HJBlearncost} as well as $\ell_{i,\textsf{H}}(t)$ that contains

\vspace{-10pt}\small\begin{align}
\partial_t\hat{\psi}_i(t) =\(\frac{\text{d}s_i(t)}{\text{d}t}\)^\T \! \nabla \hat{\psi}_i(t).
\end{align}\normalsize
According to \eqref{Eq:StateDyn}, this term introduces $\text{d}W_i(t)/\text{d}t$ that is computationally intractable. Instead, following \cite{b2}, we apply the nominal state dynamics without random wind perturbations $\text{d} s_i(t)\!=\! (\!A s_i(t)\!+\!B (a_i(t) \!+\! c_0 v_o ) ) \text{d}t$ when calculating~${\text{d}s_i(t)}/{\text{d}t}$.

\subsection{MFG Control Learning} 
In a similar vein to Algorithm~3, MFG learning control exploits ML to approximate the solutions of the HJB $\mathsf{H}\big(\psi_i^*\!(t);s_i(t),m(t) \big)\!=\!0$ and the FPK~$\mathsf{F}\!\big(m(t); s_i(t), \psi_i^*\!(t) \big)\!\!=\!0$ induced by MFG control in Sec.~\ref{Sec:MFGcontrol} as the minima of $\mathsf{H}\big(\hat{\psi}_i\!(t);s_i(t),\hat{m}(t) \big)$ and $\mathsf{F}\!\big(\hat{m}(t); s_i(t), \hat{\psi}_i\!(t) \big)$. To this end, each UAV constructs two separate ML models: the HJB model used in Algorithm~3 and an \emph{FPK model}, minimizing ${\mathsf{H}}\big(\hat{\psi}_i\!(t);s_i(t),\hat{m}(t) \big)$ (see \eqref{Eq:HJBmf} with $\hat{\psi}_i\!(t)$ and $\hat{m}(t)$) and ${\mathsf{F}}\!\big(\hat{m}(t); s_i(t), \hat{\psi}_i\!(t) \big)$ (see \eqref{Eq:FPKmodified} with $\hat{\psi}_i\!(t)$ and $\hat{m}(t)$), respectively. The FPK model has the same structure with $M_\textsf{F}$ hidden nodes, and produces the approximated MF distribution $\hat{m}(t)$ by adjusting its weight vector $w_{i,\textsf{F}}(t)$, i.e.,

\vspace{-10pt}\small\begin{align}
\hat{m}(t) = w_{i,\textsf{F}}(t)^\T \sigma_\textsf{F}(s_i(t)).
\end{align}\normalsize
Per each observation $s_i(t)$, the FPK model is trained by adjusting $w_{i,\textsf{F}}(t)$ so as to minimize the cost function $L_{i,\textsf{F}}(t)$:

\vspace{-10pt}\small\begin{align}
L_{i,\textsf{F}}(t) = \frac{1}{2}| {\mathsf{F}}\!\big(\hat{m}(t); s_i(t), \hat{\psi}_i(t) \big) |^2.
\end{align}\normalsize
The HJB model's cost function is the same as \eqref{Eq:HJBlearncost}, except for replacing its ${\mathsf{H}}\big(\hat{\psi}_i(t);s_{N_i}\!(t)\big)$ with ${\mathsf{H}}\big(\hat{\psi}_i\!(t);s_i(t),\hat{m}(t) \big)$. At time~$t$, UAV $\uav_i$'s MFG learning control is described as Algorithm 4 on the next page.

\begin{figure*}[t]
	\centering
	\setlength\abovecaptionskip{-0.0\baselineskip}
	\subfigure
	{\hspace{-13pt}\includegraphics[width = 1.1\textwidth]{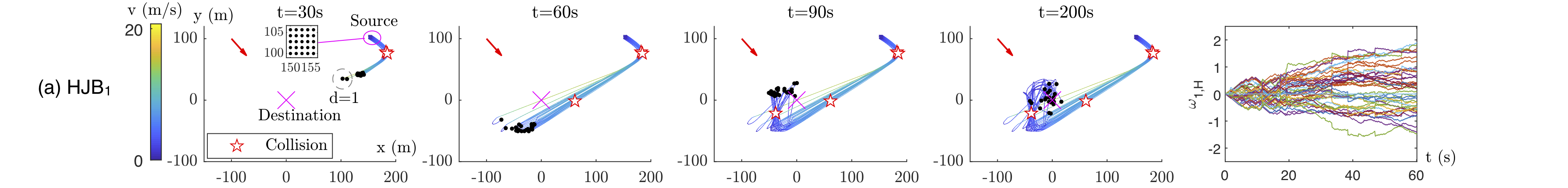}}\vskip -5pt
	\subfigure
	{\hspace{-13pt}\includegraphics[width = 1.1\textwidth]{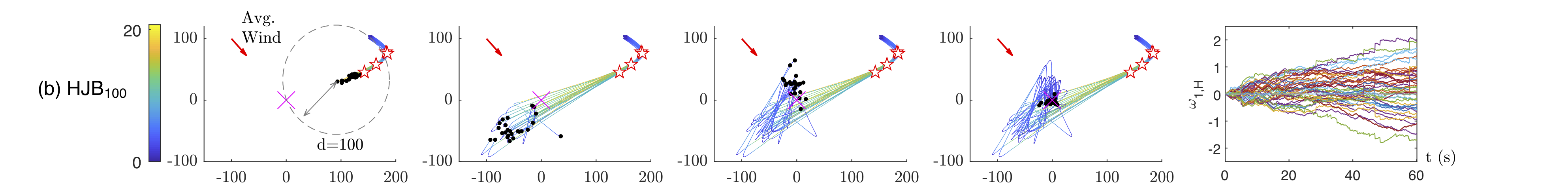}}\vskip -5pt
	\subfigure
	{\hspace{-13pt}\includegraphics[width = 1.1\textwidth]{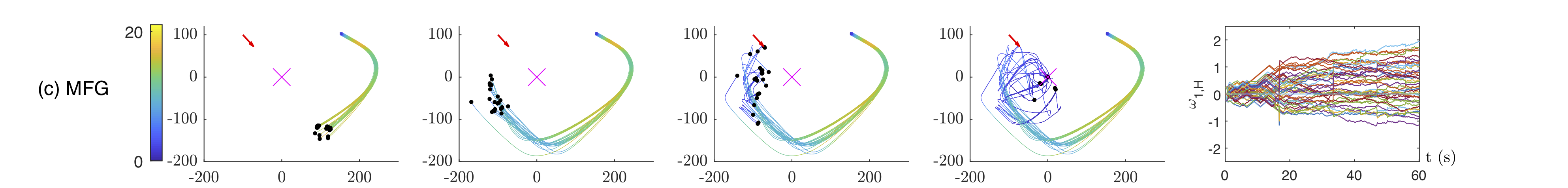}}
	\caption{Trajectory snapshots (left, $4$ subplots for each control method) of 25 UAVs under (a) $\textsf{HJB}_1$: HJB learning control with the communication range $d=1\text{m}$ (i.e., $P = 10^{-3}\text{mW} $), (b) $\textsf{HJB}_\text{100}$: HJB learning control with $d=100\text{m}$, (i.e., $ P = 10\text{mW} $), and (c) $\textsf{MFG}$: MFG learning control. During the travel time $t=0\!\sim\!200$s, $\textsf{MFG}$ shows the best flocking behavior and the most stable HJB model parameters $w_{1,\textsf{H}}$ (rightmost subplot for each control method) of a randomly selected reference UAV $\uav_1$. Consequently, $\textsf{MFG}$ yields no collision during its entire travel, in sharp contrast to $\textsf{HJB}_\text{1}$ and $\textsf{HJB}_\text{100}$.}
	\label{fig:15} \vskip -5pt
\end{figure*}

\vspace{3pt}\begin{tcolorbox}[colback=white, colframe=blue!30!black, boxrule=.5pt, top=3pt, bottom=3pt,left=10pt, right=5pt, arc=0mm]
\hspace{-8pt}\noindent\textbf{Algorithm 4. MFG Learning Control}

\hspace{-8pt}\noindent For $k\in[1,K]$:
\begin{enumerate}[leftmargin=8pt,label={\arabic*)}]
\item Update the weight $w_{i,\textsf{H}}^{[k+1]}(t)$ as:

\vspace{-10pt}\small\begin{align}
\hspace{-5pt} w_{i,\textsf{H}}^{[k+1]}(t) \!=\! w_{i,\textsf{H}}^{[k]}(t) \!-\! \mu \text{sign}(\nabla_{\!w} \ell_{i,\textsf{H}}^{[k]}(t)) - c_\textsf{H}\nabla_{\!w}R_i^{[k]}(t).\nn
\end{align}\normalsize \vspace{-10pt}

\item Calculate the value {\small$\hat{\psi}_i^{[k]}\!(t)= w_{i,\psi}^{[k]}(t)^\T \sigma_\textsf{H}(s_i(t))$\normalsize}.

\item Update the weight $w_{i,\textsf{F}}^{[k+1]}(t)$ as:

\vspace{-10pt}\small\begin{align}
\hspace{-5pt} w_{i,\textsf{F}}^{[k+1]}(t) \!=\! w_{i,\textsf{F}}^{[k]}(t) \!-\! \mu \text{sign}\(\nabla_{\!w} L_{i,\textsf{F}}^{[k]}(t)\).\nn
\end{align}\normalsize \vspace{-10pt}

\item Obtain the MF distribution~{\small$\hat{m}^{[k]}\!(t)\!=\!w_{i,\textsf{F}}^{[k]}(t)^\T \!\sigma_{\textsf{F}}(s_i(t))$\normalsize}

\item Iterate 1-4) until $k=K$.

\item Take the optimal action $a_i^*(t) \!=\! \frac{1}{2c_3}B^\T\nabla\hat{\psi}_i^{[K]}\!(t)$.
\end{enumerate}
\vspace{-5pt}\hrulefill

\hspace{-8pt}\noindent Initial MF distribution $\hat{m}^{[0]}\!(t)$ at $k=0$:
\begin{itemize}[leftmargin=8pt]
\item If $t=0$, $\hat{m}^{[0]}\!(0)=1/N\sum_{i=1}^N \mathds{1}_{s_i(t)}$, computed by collecting the states $\bar{s}_i(0)$ from N UAVs.
\item Otherwise, $\hat{m}^{[0]}\!(t)=\hat{m}^{[K-1]}\!(t-\Delta t)$.
\end{itemize}
\end{tcolorbox}

\section{Numerical Results} \label{SE:05}
In this section, we numerically compare the performances of HJB and MFG learning controls, in terms of travel time, energy consumption, and collision avoidance. For each travel, $N$ UAVs are dispatched to the origin from the source that is a square centered at $(150,100)$ in meters. At the source, each UAV is separated $\sqrt{2}$m away from each other (see Fig.~3-a), and its velocity is solely determined by the wind dynamics with $V_o = 0.1 I $ and $v_o=(1,-1)$ in m/s. Under MFG learning control, hereafter denoted as \textsf{MFG}, all UAVs are assumed to exchange their states at the source. Under HJB learning control, before every control, each UAV exchanges its state with the UAVs within the communication range $d$ meter, henceforth referred to as $\textsf{HJB}_d$, without incurring interference via frequency division multiple access (FDMA). 

For an HJB or MFG model, following \cite{b2}, a single hidden layer model is constructed, wherein each hidden node's activation function corresponds to each non-scalar term in a polynomial expansion. The polynomial is heuristically chosen as: $(1+ x_i(t) + v_{x,i}(t))^6 + (1+ y_i(t) + v_{y,i}(t))^6$ for $\sigma_\textsf{H}(s_i(t))$ and $(1 + x_i(t) + v_{x,i}(t) + y_i(t) + v_{y,i}(t))^4$ for $\sigma_\textsf{F}(s_i(t))$, where $r_i(t) = [x_i(t),y_i(t)]^{\T} $ and $ v_i(t) = [v_{x,i}(t),v_{y,i}(t)]^{\T}$. Compared to sigmoidal activations, polynomial activations enables smaller model sizes (i.e., $M_\textsf{H}\!=\!  54$, $M_\textsf{F}\!=\!  69$), yet the models are known to be less robust against unseen state observations. Optimizing the model architecture is an interesting topic for future research. Other simulation parameters are summarized as follows: \small{$ \Delta t = 1\text{s} $, $\alpha=2$, $\sigma^2=2\time10^{-2}$mW, $\theta= -10$dB, $c_0=0.1 $, $ c_1 = 100 $, $ c_2 = c_3 = 1.5 $, $ c_4 = 0.5 $, $ c_{\hspace{.5pt}\textsf{H}} = 0.5 $, $\varepsilon= 0.001$, $\mu = 0.01$, and $w_{i,\textsf{H}}(0)=w_{i,\textsf{F}}(0)=0$.}\normalsize

\figurename{ \ref{fig:15}} visualizes the trajectories of $25$ UAVs under $\textsf{HJB}_{1}$, $\textsf{HJB}_\text{100}$, and $\textsf{MFG}$. During the entire travel, UAVs under $\textsf{HJB}_{1}$ hardly communicate with each other. This makes their trajectories almost identical, causing frequent collision, where a collision is counted for an inter-UAV distance less than $0.1$m. Focusing on $\textsf{HJB}_\text{100}$, and $\textsf{MFG}$, at the beginning, all UAVs tend to follow the average wind direction to save motion energy, and then turn towards the destination. At this north-eastern turning point, $\textsf{HJB}_\text{100}$ fails to avoid collision due to its less trained HJB model. By contrast, \textsf{MFG} incurs no collision thanks to the locally iterated training operations between the HJB and FPK models (see $K$ iterations in Algorithm 4), yielding its more trained HJB (i.e., less variance in weight parameters), as observed in the rightmost subplot of Fig.~3-c. After the turning point, there is a long-distance flight of a UAV fleet. \textsf{MFG} shows the highest flight velocity owing to its better flocking, which partly compensates the longer travel distance for guaranteeing collision avoidance. Finally, at the last part of the travel, UAVs tend to hover around the destination in order to stop their movement while reaching the destination (i.e., $v_i(T)=r_i(T)=0$), which is detailed next.

Fig.~4 illustrates the accumulated number of regularizer $R_i(t)$ activations (see the details in Sec. IV-A) in the HJB models of $\textsf{HJB}_{1}$, $\textsf{HJB}_\text{100}$, and $\textsf{MFG}$ as time elapses. For all controls, $R_i(t)$ is more frequently activated near the destination (i.e., $t\geq 100$s) so as to reduce the velocity, thereby avoiding excessive hovering around and/or passing by the destination. Note that a better flocking behavior (i.e., lower inter-UAV velocities without collision) enables a more stable control without the regularization. For this reason, $\textsf{MFG}$ achieving the best flocking behavior shows the least number of $R_i(t)$ activations. With more UAVs, $\textsf{MFG}$ yields less frequent $R_i(t)$ activations. This is because the MF approximation (see Sec.~\ref{Sec:MFGcontrol}) becomes more accurate as the number of UAVs increases, providing its better flocking behavior earlier.

\begin{figure}[t]
	\centering
	\setlength\abovecaptionskip{-0.0\baselineskip}
	\hspace{-11pt}\includegraphics[trim=0cm 0cm 0cm 0.5cm,clip=true,scale=0.5]{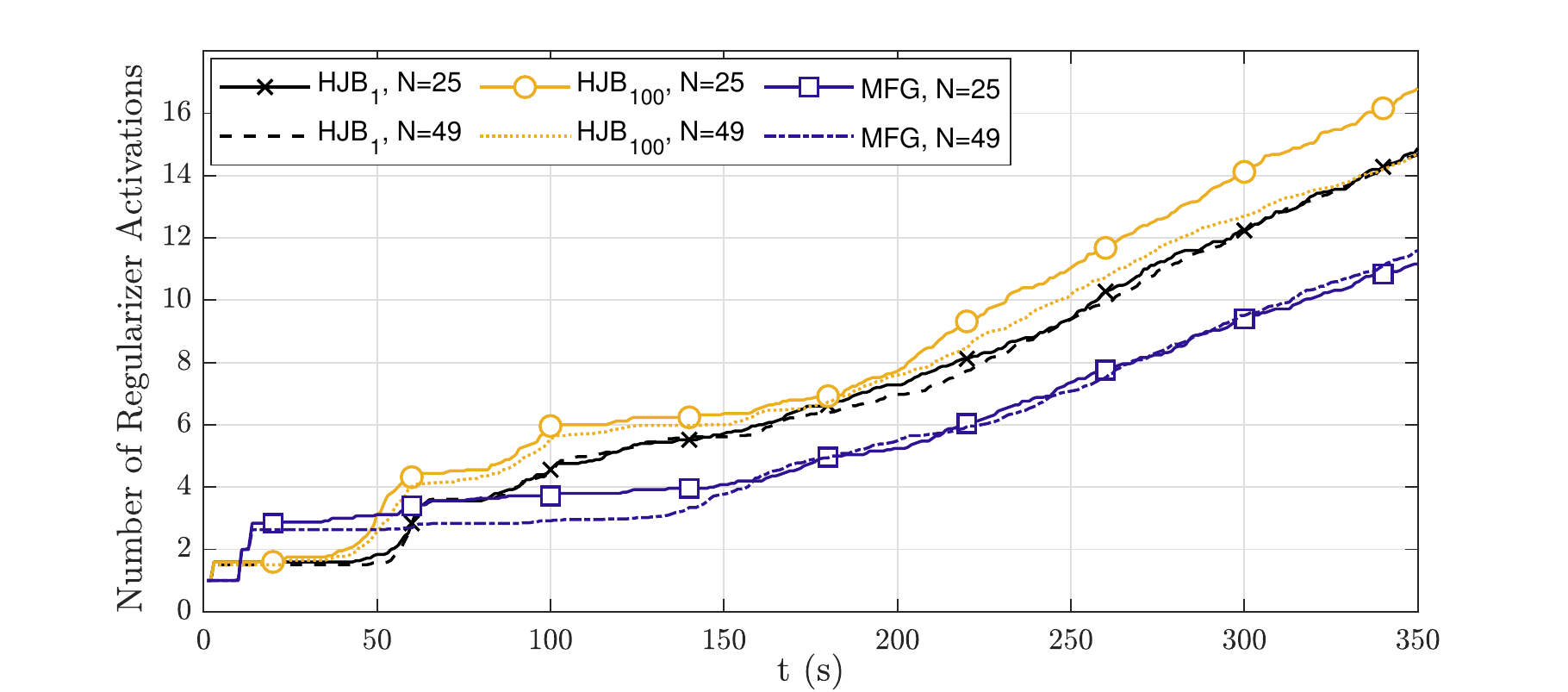}
	\caption{Accumulated number of regularizer $R_i(t)$ activations over time under $\textsf{HJB}_1$, $\textsf{HJB}_\text{100}$, and $\textsf{MFG}$ ($N=\{25, 49\}$).} \vskip -5pt

	\label{fig:17}
\end{figure}

\begin{figure}[t]
	\centering
	\setlength\abovecaptionskip{-0.0\baselineskip}
	\hspace{-11pt}\includegraphics[trim=0cm 0cm 0cm 0.3cm,clip=true,scale=0.5]{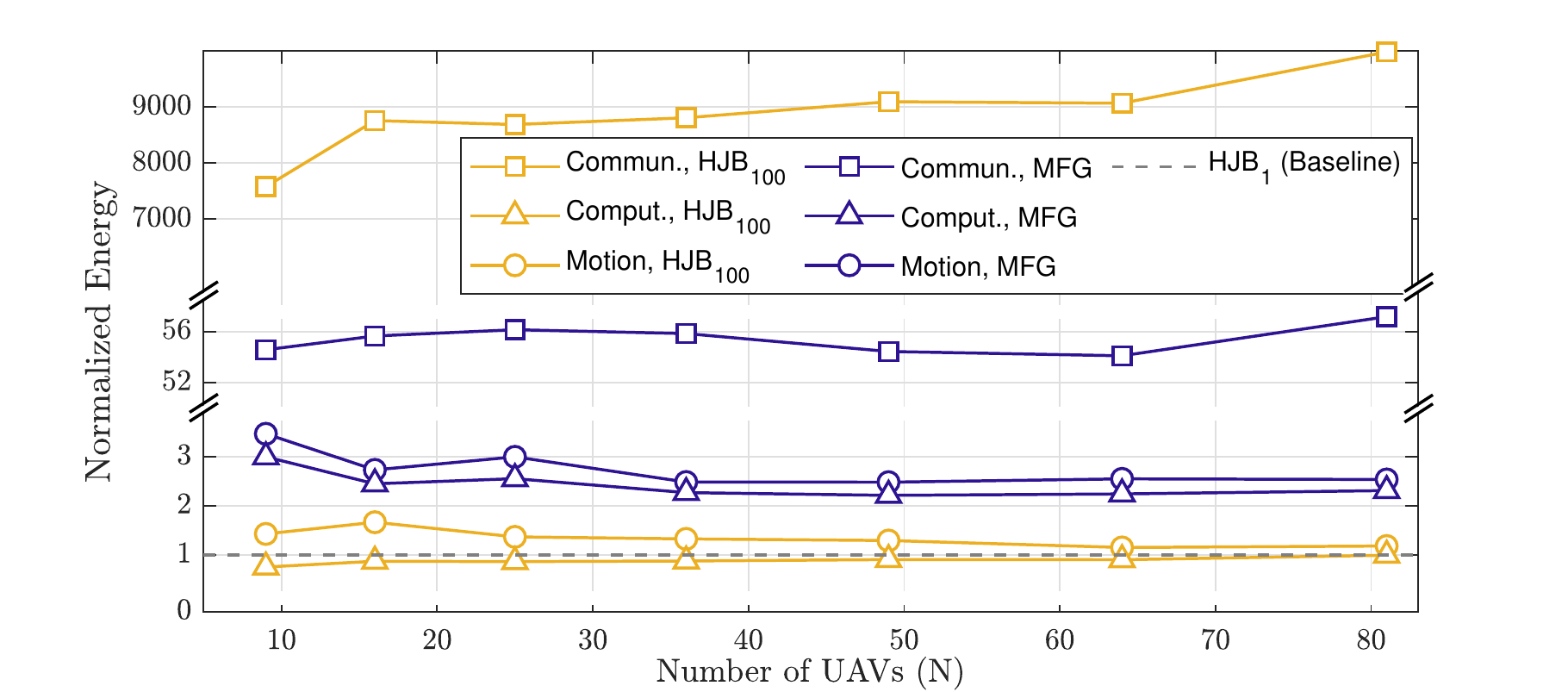}
	\caption{Comparison of communication, computation, and motion energy between $\textsf{HJB}_\text{100}$ and $\textsf{MFG}$, where each energy is normalized by the energy of $\textsf{HJB}_1$ ($N = \{9, 16, 25, 36, 64, 81\}$).}\vskip -10pt
	\label{fig:16}
\end{figure}

Lastly, Fig.~5 compares the communication, computation, and motion energy consumptions of $\textsf{HJB}_\text{100}$ and $\textsf{MFG}$ during the entire travel. Each energy is averaged over UAVs, and is normalized by the energy of $\textsf{HJB}_{1}$. We consider that communication, computation, and motion energy consumptions are proportional to the number of state exchanges, the number of gradient calculations, and $\|v_i(t)\|^2 \!+\! \| a_i(t) \|^2$, respectively. Focusing on communication energy, $\textsf{MFG}$ exchanges UAV states only once at the source, whereas $\textsf{HJB}_\text{100}$ does it for every observation. Therefore, $\textsf{MFG}$ consumes significantly less energy, irrespective of the number of UAVs, as opposed to $\textsf{HJB}_\text{100}$ whose energy increases with the number of communicating UAVs. Next, motion energy is proportional to the travel distance. As $\textsf{MFG}$ yields its longer travel distance for avoiding collision, it consumes more motion energy. For computation energy, it is also proportional to the travel distance under online learning. Besides, in contrast to $\textsf{HJB}_\text{100}$ having only an HJB model, $\textsf{MFG}$ performs gradient calculations for both HJB and FPK models, which makes $\textsf{MFG}$ consume more computation energy.

\section{conclusion} \label{SE:06}
To control massive autonomous UAVs, in this work we proposed MFG learning control algorithm that enables each UAV's real-time acceleration control in a distributed manner, by training and exploiting HJB and FPK ML models in an online way. Our simulation validated that MFG learning control guarantees collision avoidance with low communication energy, at the cost of a slight increase in computation and motion energy, compared to a baseline scheme, HJB learning control. The effectiveness of MFG learning control hinges on the level of the HJB and FPK model training. Collaborative HJB and FPK model training across UAVs via federated learning frameworks~\cite{Park:2018aa} could thus be an interesting topic for future work.

\bibliographystyle{IEEEtran}


\end{document}